# Distributed U-net model and Image Segmentation for Lung Cancer Detection

**Tianzuo Hu**

The Department of Statistics, University of California, Davis, 95616, USA

tzuhu@ucdavis.edu

**Abstract:** Until now, in the wake of the COVID-19 pandemic in 2019, lung diseases, especially diseases such as lung cancer and chronic obstructive pulmonary disease (COPD), have become an urgent global health issue. In order to mitigate the goal problem, early detection and accurate diagnosis of these conditions are critical for effective treatment and improved patient outcomes. To further research and reduce the error rate of hospital diagnoses, this comprehensive study explored the potential of computer-aided design (CAD) systems, especially utilizing advanced deep learning models such as U-Net. And compared with the literature content of other authors, this study explores the capabilities of U-Net in detail, and enhances the ability to simulate CAD systems through the VGG16 algorithm. An extensive dataset consisting of lung CT images and corresponding segmentation masks, curated collaboratively by multiple academic institutions, serves as the basis for empirical validation. In this paper, the efficiency of U-Net model is evaluated rigorously and precisely under multiple hardware configurations, such as single CPU, single GPU, distributed GPU and federated learning, and the effectiveness and development of the method in the segmentation task of lung disease are demonstrated. Empirical results clearly affirm the robust performance of the U-Net model, most effectively utilizing four GPUs for distributed learning, and these results highlight the potential of U-Net-based CAD systems for accurate and timely lung disease detection and diagnosis huge potential.

## 1.Introduction

In the post-COVID-19 era, numerous former smokers have experienced lung damage, leading to conditions such as Chronic Obstructive Pulmonary Disease and lung cancer. The Centers for Disease Control's latest available cancer incidence data is up to 2020. In the United States, there were 197,453 reported cases of new lung and bronchus cancer, resulting in 136,084 deaths from these cancers. There were 47 new cases of lung and bronchus cancer per 100,000 people, with 32 individuals succumbing to this cancer out of every 100,000 [1]. which indicates the severity of the lung cancer. In this situation, early detection plays a pivotal role in improving the prospects of successful treatment for the medical industry. Currently, most hospitals in the US mainly use low-dose computed tomography as the main screening method. However, existing diagnostic techniques include Physical Examinations, X-ray Evaluations, Magnetic Resonance Imaging, and tissue sample collection, which are often time-consuming and costly. While a range of approaches such as surgery, radiation, and chemotherapy offer potential remedies, their adverse effects on the body are substantial, and so are the challenges of prevention, diagnosis, and focused treatment to address lung cancer incidence.

Despite the availability of medical technology and dedicated medical teams to detect lung cancer, their usefulness has been very limited, nodules and tumours are often detected after they have progressed, and analysing the exact cause in a sea of images and symptoms can easily be mistaken. In order to accurately predict and identify the size, shape, and position of human organs, image segmentation algorithms in machine learning are essential. For any patient, even a slight drop in false positives can prevent many problems from occurring. Therefore, whether as a core tool or an auxiliary tool, Computer-Aided Design (CAD) technology can effectively improve diagnostic accuracy. In complex early cancer detection, CAD systems enhance the detection of lung cancer with the support of various deep-learning models [2].

For instance, Kumar et al--employed the U-Net architecture to train the pre-processed image, use image-cutting technology, and finally calculate the performance of the model. Utilizing the Adam optimizer, the U-Net architecture achieved impressive results for the CAD system. The segmentation accuracy recorded an impressive number, which is 0.9871, with a Dice Similarity Coefficient (DSC) of around 0.8205 and an Intersection/Union (IOU) score of approximately 0.7539. These findings establish the utility of U-Net in lung nodule image segmentation, as highlighted in this study [3]. Furthermore, Chon et al. implemented computer-oriented research at Stanford University. Their CAD system has three main stages, from segmentation, and nodule candidate detection to malignant tumour classification. A state-of-the-art CAD system predicts malignancies from CT scans with an AUC of 0.83. A U-Net model was trained for nodule candidate detection and applied to 3D CNN model making. The linear result is 0.665, the Vanilla 3D CNN result is 0.705, and the 3D Google net result is 0.751 [4].

Considering many factors and difficulties, this study decided to use the U-net architecture and VGG algorithm to simulate CAD because it has several significant benefits: Limited data requirements: U-net performs well even with limited training data. This is an excellent solution to the limitation of insufficient training data in our project. U-Net architecture has a symmetrical encoder-decoder structure. This simplicity facilitates faster experimentation and model tuning. Because integrating data across hospitals creates communication and security issues, and patient information is very sensitive [5], this study decided to use a distributed architecture and compare the efficiency of different Graphics Processing Units (GPUs) against the model, hoping to find a dynamic link between model performance and the protection of sensitive medical data.

## 2. Methods

*2.1 Dataset preparation*

In a collaborative effort between Qatar and Dhaka Universities, with additional contributions from collaborators in Pakistan and Malaysia, this study utilized a substantial dataset comprising chest X-ray images [6]. Remarkably extensive, the dataset encompasses 11,956 cases of COVID-19, 11,263 cases of non-COVID infections, and 10,701 normal cases. Each dataset entry is meticulously paired with a lung segmentation mask, furnishing comprehensive annotations. A significant inclusion is the specialized subset of 2,913 masks for COVID-19 identification—an extension of the QaTaCov project [6]. For experimentation, a meticulously curated subset of 1,000 lung CT images has been employed in this research, aimed at lung image identification amidst infections. Some sample images can be found in Figure 1.

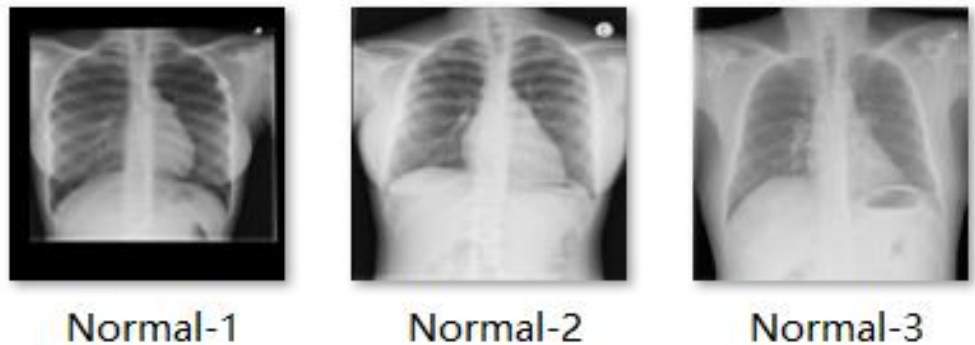

**Figure 1.** The top three of one thousand CT image of human lung shows the impact of COVID-19.

**Table 1.** This is the hardware environment of distributed U-Net model in the experiment.

| GPU: RTX 2080 Ti |
|---|

**Table 2.** Here is the comparison environment of U-Net model that using CPU during the experiment.

| CPU: Intel Core i7-10870H |
|---|

*2.2 Distributed and Federal Synchronous Stochastic Gradient Descent (SGD)*

In this study, the U-Net neural network architecture was specifically designed for semantic segmentation tasks, offering the flexibility to incorporate either a VGG16 or ResNet-50 backbone. At its core, the architecture features the U-Net Up module shown in Figure 2, which employs convolutional layers, up-sampling operations, and Re-LU activations to enhance feature representation [9].

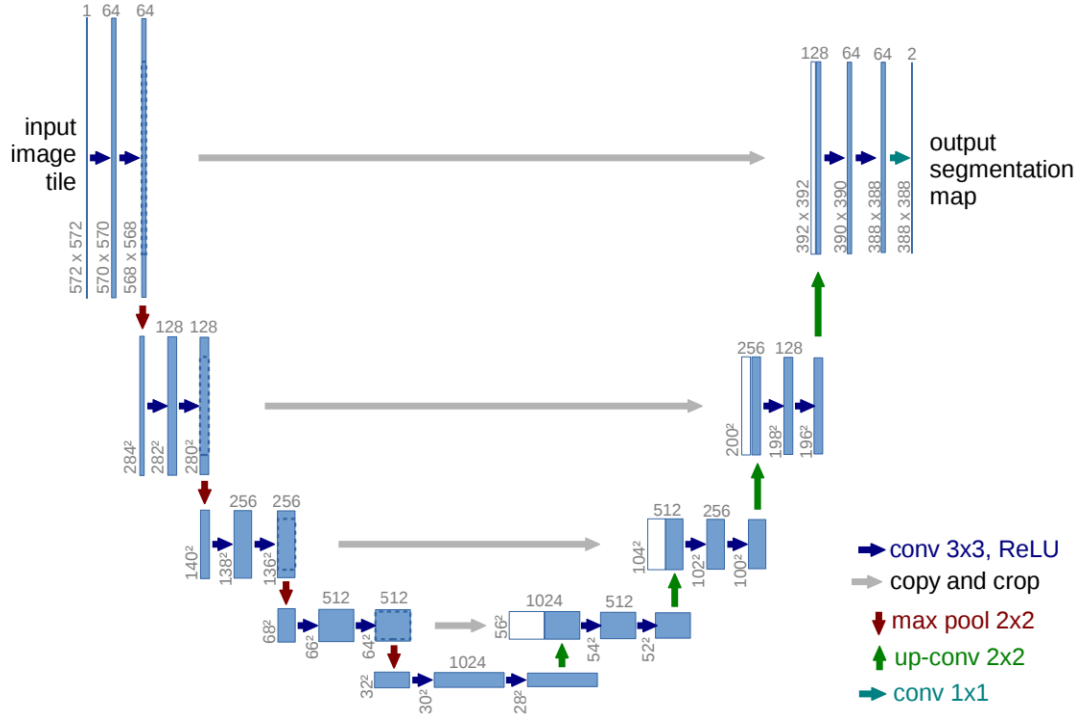

**Figure 2.** The U-Net model's visualization architecture is depicted here. The white boxes indicate duplicated feature maps, the blue blocks represent feature maps with multiple channels, and their channel count is shown on top. On the bottom left, you'll find the size information. And all arrows can direct to each following steps [10].

Notably, the architecture offers controlled training adaptability through methods like freeze backbone and unfreeze backbone, enabling fine-tuning for specific components. In summary, the proposed U-net architecture stands as a potent solution for image segmentation:

Limited Data Requirements: U-Net can perform well even with limited training data. This perfectly solves the limitation of insufficient training data volume in our project, not really rely on the equipment.

Architectural Simplicity: U-Net architecture has a symmetric encoder-decoder structure, which makes it relatively straightforward to implement and understand. This simplicity can facilitate quicker experimentation and model adjustments.

The predict section is an integration of the U-Net model, a dedicated deep learning architecture for image segmentation tasks, into a versatile and multifunctional framework. Serving as the core of the script, the U-net class encapsulates U-Net's essential components, including its encoder, decoder, and skip connections. Through distinct methods within the U-net class, the script facilitates diverse operations such as single-image prediction, real-time video analysis, FPS evaluation, and batch processing. This integration enhances the model's utility by accommodating various operational modes.

*2.3 Implementation details*

The experiment involves training the U-Net model on lung images using different approaches and evaluating key performance indicators. From the equation (1), K: the number of client devices participating in the training. $F_k(w)$: the loss function of the kth client.

$$f(w) = \sum_{k=1}^{k} \frac{n_k}{n} F_k(w)$$

**Figure 3.** This is the loss expression for Federated Learning.

Observed from Figure 3, the U-Net model consists of three main components. The first component focuses on backbone feature extraction, similar to the VGG architecture. By applying convolution and maximum pooling layers, five preliminary feature layers are obtained from the backbone. The second component enhances feature extraction by employing up-sampling and feature fusion techniques on the five preliminary feature layers to create a final integrated feature layer. The third component involves utilizing the last effective feature layer for pixel-wise classification.

**Table 3.** Whole experiment contains 3 main training parts, single 2080 GPU for SGD algorithm, 4 GPUs for DSGD algorithm [11], 4 2080 GPUs for distributed learning, and 4 GPUs for Federal Learning which can show their differences at last [12].

| Methods | Training Arrangement |
|---------|----------------------|
| SGD     | 1000 data on 1 GPU   |
| DSGD    | 1000 data on 4 GPUs  |
| FL      | 4 GPUs (each GPU has 250 data) |

During the training process, several critical metrics are monitored. The total loss value assesses the convergence of the model's predictions to the actual segmentation labels in each iteration. Lower loss values indicate improved convergence.

The F-score metric, which combines accuracy and recall, serves as a comprehensive measure of segmentation performance. The learning rate start from 0.00001, which is dynamically adjusted through cosine annealing to optimize the step size for better convergence during training. To enhance convergence efficiency, a freeze and thaw strategy is employed, involving alternating cycles of freezing specific layer weights and training the entire model.

$$F - score = \left(\frac{Precision^{-1} + Recall^{-1}}{2}\right)^{-1}$$

**Figure 4.** The formula to calculate the F-score metric.

During the second phase of training on a single CPU, it comprises 10 cycles. At the conclusion of each cycle, the model's performance is assessed using a validation set, and the weights of the best-performing model are preserved.

**3. Results and discussion**

*3.1 Training result*

For the main training, the same learning rate, update strategy and optimizer are used for all training sessions, 10 epochs are trained for each training section, and transfer learning methods are used to reduce the hardware requirements of the training process. The first five epochs freeze the backbone weights, and the last five epochs train the entire model. By comparison to

GPUs, in terms of experimental conclusions, the Figure 5 shows that U-Net model performs well in lung disease segmentation tasks, achieving low loss, the best the method should be 4 GPUs- 8 batch, which is close to 0.02 and the lowest loss in different method. In addition, training strategy and the application of VGG as a feature extraction network also have a positive impact on optimizing training efficiency and improving performance.

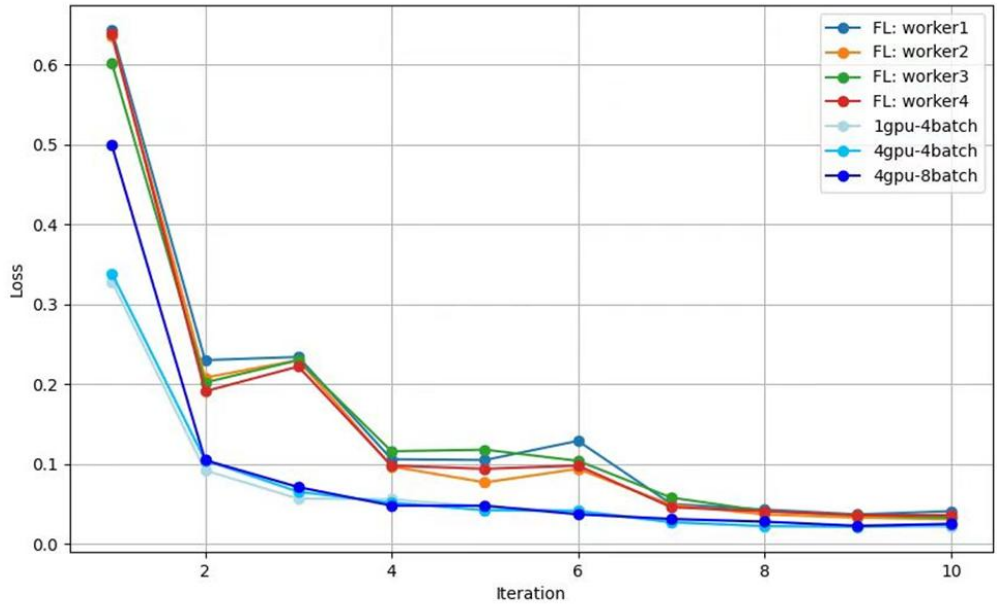

**Figure 5.** The final result is visualized in a curve graph. On the horizontal axis, the graph shows the number of iterations, while the vertical axis represents the parameter that varies across different U-Net models as they learn.

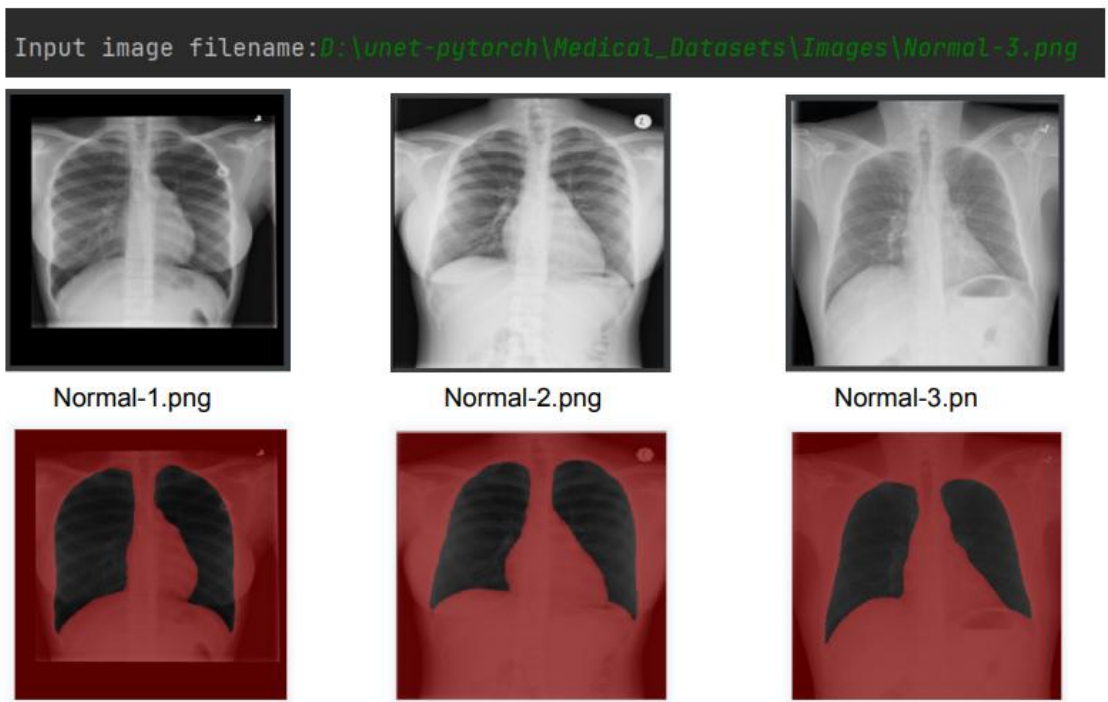

**Figure 6.** The results of human lung CT images can be recognized by distributed U-Net model, and the first three images in the database are taken as examples to prove the accuracy of recognition.

**Table 4.** Every epoch loss of the model on the second part of 1 CPU SGD training, the result of training loss is 0.019, it's the lowest but takes very long time- 12 hours and 37 minutes, which means it's not efficient.

| Steps | Training loss of U-Net model on CPU | F-scores | Learning rates |
| --- | --- | --- | --- |
| 1 | 0.243 | 0.783 | 1e-5 |
| 2 | 0.096 | 0.942 | 0.0001 |
| 3 | 0.062 | 0.963 | 9.62e-5 |
| 4 | 0.047 | 0.971 | 8.55e-5 |
| 5 | 0.041 | 0.975 | 6.94e-5 |
| 6 | 0.050 | 0.971 | 5.05e-5 |
| 7 | 0.027 | 0.982 | 3.16e-5 |
| 8 | 0.022 | 0.986 | 1.55e-5 |
| 9 | 0.022 | 0.985 | 4.77e-6 |
| 10 | 0.019 | 0.987 | 1e-6 |

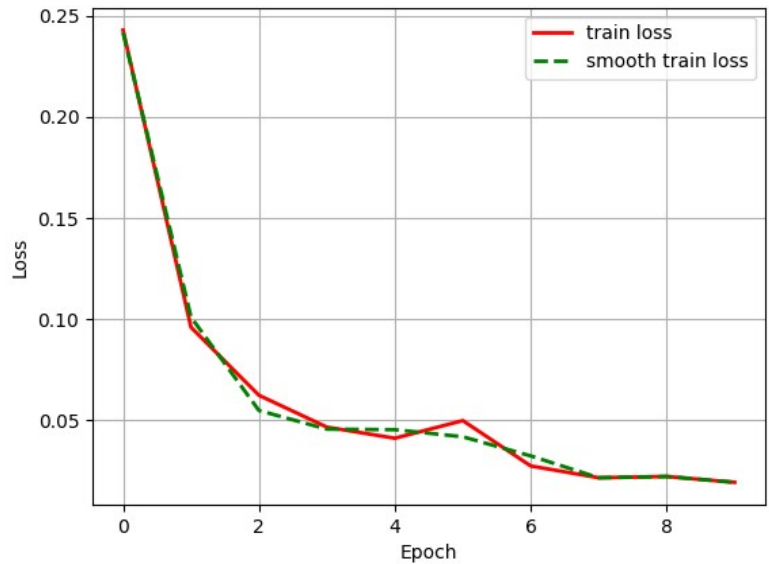

**Figure 7.** This curve graph is about 1 CPU SGD training process, the loss at the beginning is large, and then start decreasing rapidly, at last, the loss is nearly 0.019.

The depicted trajectory of the training loss curve on the Table 4 and Figure 7 denotes the stability and convergence exhibited by the model throughout its training. This correspondence aligns effectively with the conceptual framework discussed earlier. This congruence serves as an indication of the model's coherent training progression, characterized by a gradual approach towards optimal performance. As a result, the consistently smooth loss curve pattern signifies proficient optimization.

On the Table 5, regarding the temporal dimension of training, the empirical findings reveal that the utilization of GPUs in contrast to CPUs markedly expedites the training procedure. Additionally, a noticeable reduction in training time is observed when comparing Distributed Synchronous SGD to Stochastic Gradient Descent.

Analyzing the visual representation provided below, it is evident that operations conducted through multiple GPUs in both distributed and SGD settings manifest notably heightened efficiency in simultaneous processing when juxtaposed with individual GPU and CPU scenarios.

**Table 5.** Final result and comparison of each method during the experiment, and provide the training time comparison above. The result is that the time on 4 GPU for DSGD is close to the 4 GPU for FL, but actually their training loss are different.

| Methods | Time |
| --- | --- |
| 1 CPU for SGD | 12 hours 37min 50s |
| 1 CPU for SGD | 14 min 54s |
| 4 GPU for DSGD | 5 min 28s |
| 4 GPU for FL | 5 min 28s |

*3.2 Outlook:*

There are some aspects that can do in the following improvement:

(1) A crucial focal point lies in the enhancement of communication efficiency. This entails the exploration and advancement of innovative techniques encompassing compression, quantization, and optimization algorithms within the network. The overarching goal is the reduction of transmitted data volume between edge devices and the central server throughout the federated learning process.

(2) Another pivotal consideration is the accommodation of non-identically and non-independently distributed (non-IID) data scenarios. A common occurrence involves the receipt of data volumes varying across different hospitals. The integration of non-IID principles into the project becomes imperative to align with real-world conditions and improve the efficacy of the solution.

(3) The potential of mix-up data augmentation technology holds substantial promise. Specifically, within 2D image datasets, instances of object overlap may occur, potentially leading to elevated false positive rates, as seen in lung nodule detection. By incorporating the mix-up technique, prospects for mitigating this challenge and refining accuracy emerge.

**4. Conclusion:**

In conclusion, this research highlights the enormous potential of machine learning, specifically the U-Net architecture, to enhance the precision as well as effectiveness of lung cancer segmentation from CT scans. The model obtained considerable reductions in training time compared to CPU-based systems by utilizing distributed and federal learning methods across numerous GPUs. This not only speeds up the diagnostic process, but it also provides a scalable approach for dealing with huge datasets, which is critical in real-world clinical situations. Our results show that the U-Net model performed remarkably accurate when trained on a dataset of 1,000 lung CT images, with a loss value close to 0.021. The observed efficiency ranking was as follows: 4 GPUs versus 4 GPUs Federal Learning > 1 GPU > 1 CPU, which means the distributed U-Net model can accurately achieve the result. Although 1 CPU can get lower loss, it still took a longer time. At last, this study highlights avenues for further improvement, including enhancing communication efficiency, addressing non-IID data scenarios, and exploring mix-up data augmentation techniques. These refinements hold promise for advancing the field of lung cancer detection and improving patient outcomes.